\documentclass{elsart}
\usepackage{amssymb}
\usepackage{amsmath}
\usepackage{graphicx}

\begin{document}

\begin{frontmatter}



\title{Characterization and modelling of the hollow beam produced by a
real conical lens}


\author{Beno\^{\i}t D\'{e}pret}
\author{Philippe Verkerk}
\author{Daniel Hennequin}
\address{Laboratoire de Physique des Lasers, Atomes et Mol\'{e}cules, UMR
CNRS, Centre d'Etudes et de Recherches Lasers et Applications, Universit\'{e}
des Sciences et Technologies de Lille, F-59655 Villeneuve d'Ascq cedex -
France}

\date{\today }

\begin{abstract}
The properties of the hollow beam produced by a conical lens are studied in
detail. In particular, the impact of a rounded vertex is examined. It is
shown that it could lead to drastic changes in the transverse distribution
of the hollow beam, determined by the ratio between the transverse size of
the incident beam and the size of the blunt area. An adequate choice for
this ratio allows us to either minimize the losses or optimize the
distribution symmetry.
\end{abstract}

\end{frontmatter}


\section{Introduction}

\begin{figure}[tph]
\centerline{\includegraphics[width=10cm]{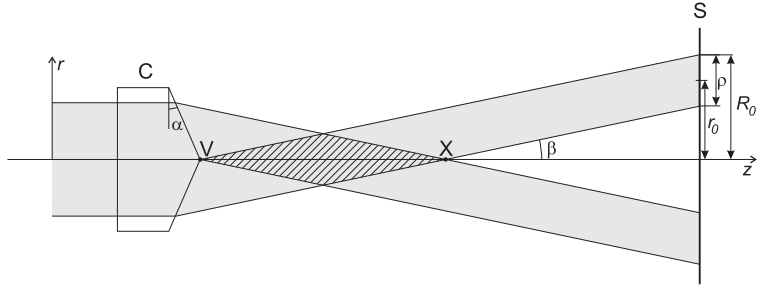}}
\caption{Schematic behavior, in the geometrical approximation, of a
collimated beam through a conical lens C of base angle $\protect\alpha$. The
propagation axis is the $z$ axis, with its origin at the axicon vertex.
After the conical lens, the beam is globally deviated towards the $z$ axis
with an angle $\protect\beta$. Point V is on the axicon vertex, while point
X corresponds to the abscissa $z_0$ where a hole appears in the center of
the beam in geometrical approximation. The hatched zone between V and X is
the Bessel zone. For $z>z_0$, the hollow beam has a radius $r_0$ and a
thickness $\protect\rho$.}
\label{fig:schema}
\end{figure}

Axicons are a family of cylindrical symmetrical optical systems that produce
a line focus rather than a point focus from incident collimated beam\cite%
{McLeod54}. There are several types of axicons, working either by reflection
or by transmission, and being either converging or diverging, but the most
common one is probably the conical lens. Such a lens is a cone made with a
material of index $n$, and a basis perpendicular to its main $z$ axis (Fig. %
\ref{fig:schema}). An incident collimated beam, propagating along the $z$
axis, is deviated towards the main axis of the lens, with an angle $\beta
=\left( n-1\right) \alpha $, where $\alpha $ is the base angle of the cone. $%
\alpha $ is usually a small angle, typically few degrees or less. Beyond the
lens, two zones must be distinguished: in the first one, just after the
lens, all deviated beams spatially coexist partially (in the hatched zone of
Fig. \ref{fig:schema}), resulting in a diffraction-free beam. Beyond this
zone, beams deviated along different directions do not overlap any more, and
the intensity becomes distributed on a ring. Such hollow beams have recently
regain interest, in the perspective of their use in optical trapping of atoms%
\cite{Manek98,Cacciapuoti2001,Kulin2001} and Bose-Einstein condensates\cite%
{Wright2000}.

These applications require specific properties, as e.g. a zero intensity in
the center of the beam, as predicted by the diffraction theory\cite%
{Belanger78}. However, in most of the experimental realizations, the
obtained pattern is more complicated: a usual observed defect is a drastic
asymmetry of the distribution with a tail towards the center, possibly
secondary rings and thus a non zero intensity at the center\cite{Manek98}.
Practically, this leads to the need of a mask to clean the inner region of
the hollow beam, and thus a loss of light power and additional diffraction
effects, which are both major inconveniences in the experimental uses of
such hollow beams. The source of these differences with respect to the
theory is not clearly identified. As the conical lens vertex plays a central
role in the transverse distribution of the hollow beam, it is natural to
search for the origin of these defects in the properties of this point, in
particular by taking into consideration its real shape. In the present
paper, we study the modifications induced by the bluntness of the conical
lens vertex on the hollow beam transverse shape. We show in particular that
the ratio between the transverse size of the incident beam and the size of
the blunt area determines the transverse distribution, and that an adequate
choice for this ratio allows us to either minimize the losses or optimize
the distribution symmetry.

\section{Geometrical propagation}

Fig. \ref{fig:schema} describes schematically the geometrical beam
propagation through a conical lens. The diffraction-free zone appears
between the conical lens vertex V, and the point X where all deviated beams
separate. The spatial distribution of the light field in this zone is the
superimposition of the interferences between deviated beams and diffraction
on the conical lens vertex V. The transverse distribution of the resulting
field follows a Bessel function\cite{Herman91}. Its exact nature depends on
the shape of the incident beam. For example, an appropriate incident
Laguerre-Gaussian beam is able to generate a high-order Bessel beam of
arbitrary order\cite{Arlt2000}. The main interest of Bessel beams is that
they propagate without diffraction, i.e. their intensity transverse profile
remains invariant upon propagation on macroscopic distances\cite{Durnin87}.
Bessel beams produced by axicons have been extensively studied, within the
framework of numerous applications, such as second-harmonic generation\cite%
{Arlt99}, optical parametric generator \cite{Gadonas1998}, and atom trapping %
\cite{Arlt2001}.

Beyond point X, i.e. for $z>z_{0}$, with $z_{0}=r_{i}/\beta $, where $r_{i}$
is the radius of the incident beam, all deviated beams separate (in the
geometrical approximation), and a hole with radius $R_{0}-\rho $ appears in
the center of the beam. $R_{0}=\beta z$ is the distance between the $z$ axis
and the ray crossing the conical lens vertex, while $\rho $ is the thickness
of the ring ($\rho =r_{i}$ in the present case). It is also practical to
define the radius $r_{0}$ of the ring, with $r_{0}=R_{0}-\rho /2$ (Fig. \ref%
{fig:schema}). Such hollow beams are fundamentally different from hollow
Laguerre-Gaussian modes. First, in the last ones, the circulation of the
phase around the pattern center is equal to $2\pi l$, where $l>0$ is the
angular index of the mode, while in the axicon hollow beam, it is equal to
that of the incident beam, i.e. zero in the case of a plane wave or a TEM$%
_{00}$ mode. This phase variation has no consequence on the intensity
transverse distribution, but is critical in field dependent systems, e.g. if
the hollow beam interferes with another beam. The second
point concerns directly the intensity distribution: the axicon hollow beam
can be realized with arbitrary ratio between the radius and the thickness of
the beam, as both quantities depend differently on $z$, while in
Laguerre-Gaussian modes, this ratio is fixed by the Laguerre polynomial
distribution and does not vary with $z$, except of course if it is transformed
by an axicon\cite{Arlt2001b}. Axicon hollow beams can be used in
different applications, from laser machining\cite{Rioux78} to optical traps
for cold atoms\cite{Manek98,Cacciapuoti2001,Kulin2001} and Bose-Einstein
condensates\cite{Wright2000}.

\section{Gaussian beam propagation}

\begin{figure}[tph]
\centerline{\includegraphics[width=10cm]{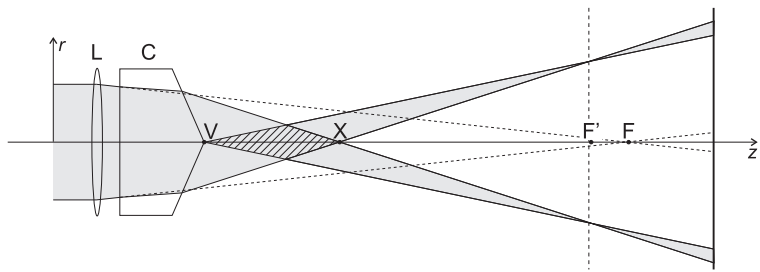}}
\caption{Schematic behavior, in the geometrical approximation, of a
collimated beam through a converging lens L followed by a conical lens C.
Point V is on the axicon vertex, while point X corresponds to the abscissa $%
z_0$ where a hole appears in the center of the beam. The hatched zone
between V and X is the Bessel zone.}
\label{fig:schema2}
\end{figure}

In practical realizations, incident beams are often produced by a laser, and
thus are gaussian beams. In that case, the incident beam cannot be
considered as collimated, but is characterized by a waist $%
w_{0}$ where the beam spot size is minimum. To control the waist location
and size, a convergent lens is usually inserted before the conical lens. In
particular, it is easy with such a doublet to locate the waist after the
conical lens. Fig. \ref{fig:schema2} shows the propagation of a collimated
incident beam through a convergent lens followed by a conical one, in the
geometrical approximation. As for collimated beams, an incident ray at the
distance $r$ from the $z$ axis, is deviated with an angle $\beta $, with $%
\beta \left( r\right) =\left( n-1\right) \alpha +r/f$, where $f$ is the
focal length of the convergent lens. We can also define a distance $z_{0}$
such that for $z>z_{0}$, a hole appears in the center of the beam. For thin
lenses, a simple calculation leads to $z_{0}=r_{i}\left( f-d\right) /\left(
r_{i}+\left( n-1\right) \alpha f\right) $ where $d$ is the distance between
the two lenses. As in the single conical lens case, $R_{0}$ and $\rho $,
which now depend both on $z$, can be easily calculated. In particular $%
R_{0}=\left( n-1\right) \alpha z$ has the same value as in the single
conical lens scheme. The point focus in F is transformed in a ring focus of
radius $R_{0}=\left( n-1\right) \alpha \left( f-d\right) $, \ at location F'$%
{}\simeq {}$F, where F is the focus point of the convergent lens (the
interval between both points is in $\alpha ^{2}$). If the gaussian
properties of the incident beam are taken into consideration, all these
quantities becomes indefinite, as the transverse expansion of a gaussian
beam is theoretically infinite. However, for large incident beams (i.e. $%
w_{i}>>\lambda $, $w_{i}$ being the incident beam size at $1/e^{2}$ and $%
\lambda $ the wavelength), the geometrical approach remains a good
approximation. Therefore, the above expressions, where $r_{i}$ is replaced
by $w_{i}$, can be used, keeping in mind that they concern $1/e^{2}$
intensity limits.

This geometrical approach is unable to describe the\
thickness $\rho $ of the ring. To evaluate it, it is necessary to take into
account the diffraction of the beam during the propagation, and in
particular on the conical lens vertex. Such a theory shows that when the
convergent conical lens doublet is illuminated by a gaussian beam, the
transverse distribution of the ring at the waist location, is also gaussian,
symmetrical with respect to the $r=R_{0}$ circle \cite{Belanger78}. Its
thickness at $1/e^{2}$ is $\rho _{0}=1.65w_{0}$, comparable to the beam
waist $w_{0}$ produced by the converging lens without axicon. The field
amplitude of the wave after the lens is given by the usual Fresnel
approximation to the Kirchhoff integral: 
\begin{equation}
u(P)=-\frac{i}{2\lambda }\int_{S}u_{i}\left( M\right) \tau \left( M\right) I%
\frac{e^{ik\zeta }}{\zeta }dS  \label{int1}
\end{equation}%
where $\lambda $ is the wavelength and $k=2\pi /\lambda $ is the wave
vector. $u\left( P\right) $ is the complex amplitude of the field in point $%
P $ of the observation plane, $u_{i}\left( M\right) $ the incident field in
point $M$ of the diffraction plane and $\tau \left( M\right) $ the
transmission function of point $M$. The integration is done on the whole
diffracting surface $S$. $\zeta $ is the optical distance between $M$ and $P$%
, and thus $e^{ik\zeta }/\zeta $ depicts the spherical wave expanding out
from point $M$. The inclination factor $I$ depicts the angle of the incident
and diffracted beams with the normal to the surface $S$. For small apertures
and paraxial beams, it is approximated by $I=-2$, so that Eq. \ref{int1}
becomes: 
\begin{equation}
u(P)=\frac{i}{\lambda }\int_{S}u_{i}\left( M\right) \tau \left( M\right) 
\frac{e^{ik\zeta }}{\zeta }dS  \label{int2}
\end{equation}%
The transmission function depends of course on the considered optical
elements. It is $\tau _{L}\left( M\right) $ and $\tau _{C}\left( M\right) $
for respectively a lens and a conical lens: 
\begin{subequations}
\label{trans}
\begin{eqnarray}
\tau _{L}\left( M\right) &=&\exp \left( -ik\frac{r^{2}}{2f}\right) \\
\tau _{C}\left( M\right) &=&\exp \left( ik\left( n-1\right) r\tan \alpha
\right)
\end{eqnarray}%
\end{subequations}
where $r$ is the distance from point $M$ to the optical axis, $f$ is the
focal length of the lens and $n$ the optical index. Eqs \ref{trans} do not
take into account a fixed phase shift induced by the thickness of the lenses.

To compute the field amplitude after the conical lens and compare it with
experimental measurements, we directly evaluate Eq. \ref{int2} through a
standard integration software, without any additional approximation.
Experimentally, we used a commercial conical lens from Bern optics, with $%
\alpha =2%
{{}^\circ}%
$ and a radius of 5 mm. It is in BK7 glass with an optical index $n=1.51$.
The incident gaussian beam, provided by a laser diode with $\lambda =852$%
~nm, has a waist $w_{i}=645$~$\mu $m, so that it can be considered as
collimated, and we have $z_{0}=w_{i}/\beta =36%
\mathop{\rm mm}%
$. The signal is detected through 2D and 1D CCD cameras, and therefore we
record the {\em intensity} of the pattern rather than its field amplitude.
Thus the comparison between experimental and numerical results is performed
through the intensity transverse distribution.

\begin{figure}[tph]
\centerline{\includegraphics[width=10cm]{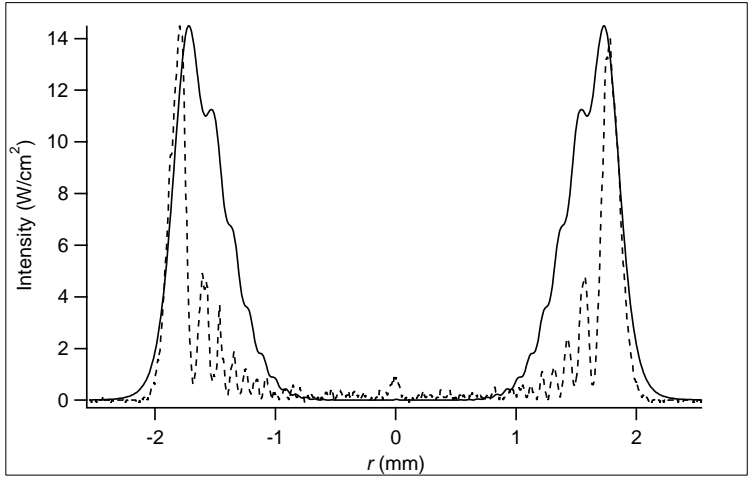}}
\caption{Experimental (dashed line) and theoretical (full line) transverse
intensity profile of the beam in $z=115\mathop{\rm mm}=3.2z_0$ after the
axicon, obtained when the incident beam, with a waist $w_i=645$~$\protect\mu 
$m, crosses a unique conical lens. The theoretical profile corresponds to
the modelization of the axicon with a ponctual vertex. The vertical scale
corresponds to the theoretical curve, when the total light power is $100$
mW. Units for the experimental curve are arbitrary.}
\label{fig:diff}
\end{figure}

Let us first illustrate the limits of the present model with a spectacular
example: we consider the transverse distribution of the experimental beam
described above, at a distance $z=115%
\mathop{\rm mm}%
=3.2\,z_{0}$ after the conical lens. Fig. \ref{fig:diff} shows the
experimental (dashed line) and theoretical (full line) intensity
distributions. While the ring diameters are identical, it is clear that the
present model is unable to reproduce the details of the distribution. In
particular, the experimental profile shows a thin main ring, with inner
contrasted secondary rings, looking like diffracting rings. On the contrary,
the model predicts a unique thick ring, slightly modulated by diffraction.
Thus it is clear that the diffracting elements are not treated correctly in
the model.

\section{Modelling of a real axicon}

\begin{figure}[tph]
\centerline{\includegraphics[width=10cm]{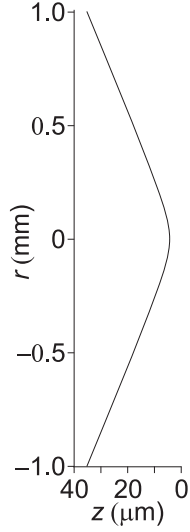}}
\caption{Schematic representation of the axicon shape taking into account
the flatness of its vertex. All parameters are those of Fig. \ref{fig:diff},
with $R_{\exp }=3.5$~mm. The origin of $z$ is the conical lens vertex.}
\label{fig:hyperb}
\end{figure}

\begin{figure}[tph]
\centerline{\includegraphics[width=10cm]{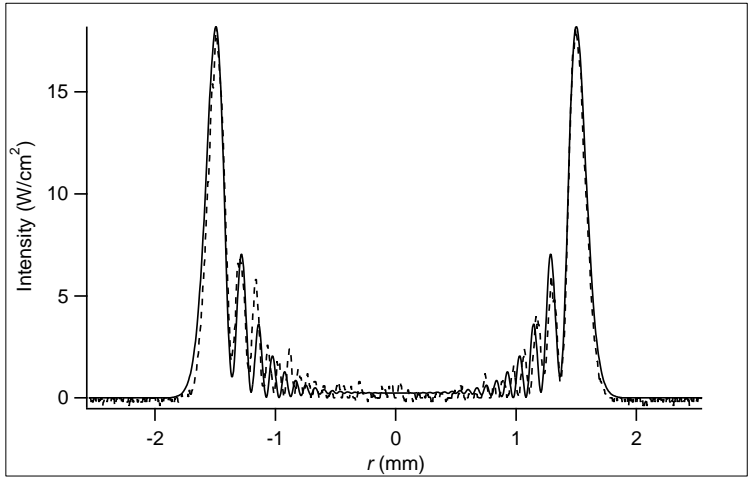}}
\caption{Same as Fig. \ref{fig:diff}, except that the theoretical profile
corresponds to the modelization of the axicon with the shape of Fig. \ref%
{fig:hyperb}.}
\label{fig:fit}
\end{figure}

\begin{figure}[tph]
\centerline{\includegraphics[width=10cm]{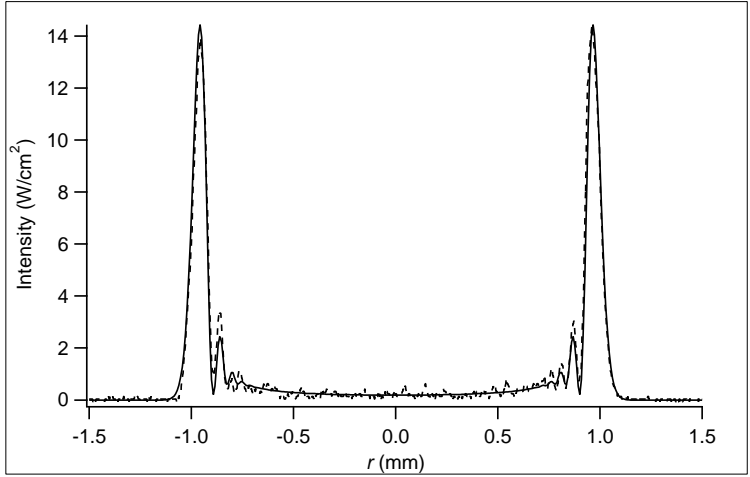}}
\caption{Experimental (dashed line) and theoretical (full line) transverse
intensity profile of the beam in $z=65\mathop{\rm mm}=2.7z_0$ after the
axicon, obtained when the incident beam, with a waist $w_i=645$~$\protect\mu 
$m, crosses a convergent conical lens doublet. The convergent lens has a
focal length $f=100$~mm, and is located at $z=-10$~mm. The theoretical
profile corresponds to the modelization of the axicon with a rounded vertex.
Vertical scales as in Fig. \ref{fig:diff}.}
\label{fig:fitdoublet}
\end{figure}

To enhance the model, we take into consideration the bluntness of the
conical lens vertex, by introducing an hyperbolic correction to the conical
lens shape (fig. \ref{fig:hyperb}). We introduce the radius of curvature $R$
of the hyperbole, so that the thickness $e\left( r\right) $ of the lens is: 
\begin{equation}
e\left( r\right) =e_{0}-R\tan ^{2}\left( \alpha \right) \sqrt{1+\frac{r^{2}}{%
R^{2}\tan ^{2}\left( \alpha \right) }}
\end{equation}%
where $e_{0}$ is the thickness of the lens for $R=0$, i.e. when the vertex
is a point. To evaluate the radius of curvature $R_{\exp }$ of our actual
conical lens, we use a fit to adjust experimental and theoretical intensity
distributions. The best fit, corresponding to $R_{\exp }=3.5$~mm, leads to a
satisfying concordance between experiments and theory (Fig. \ref{fig:fit}).
The remaining differences between both curves concern the amplitude of the
secondary rings, but not their number neither their location. We have also
applied our model to non collimated beams, i.e. when the conical lens is
used with a convergent lens. As shown on Fig. \ref{fig:fitdoublet}, the
transverse distribution of the beam after the doublet is well reproduced.
Therefore\ the hyperbolic approximation of the vertex appears to be good
enough to predict and optimize the use of an axicon to produce a hollow beam.

\begin{figure}[tph]
\centerline{\includegraphics[width=10cm]{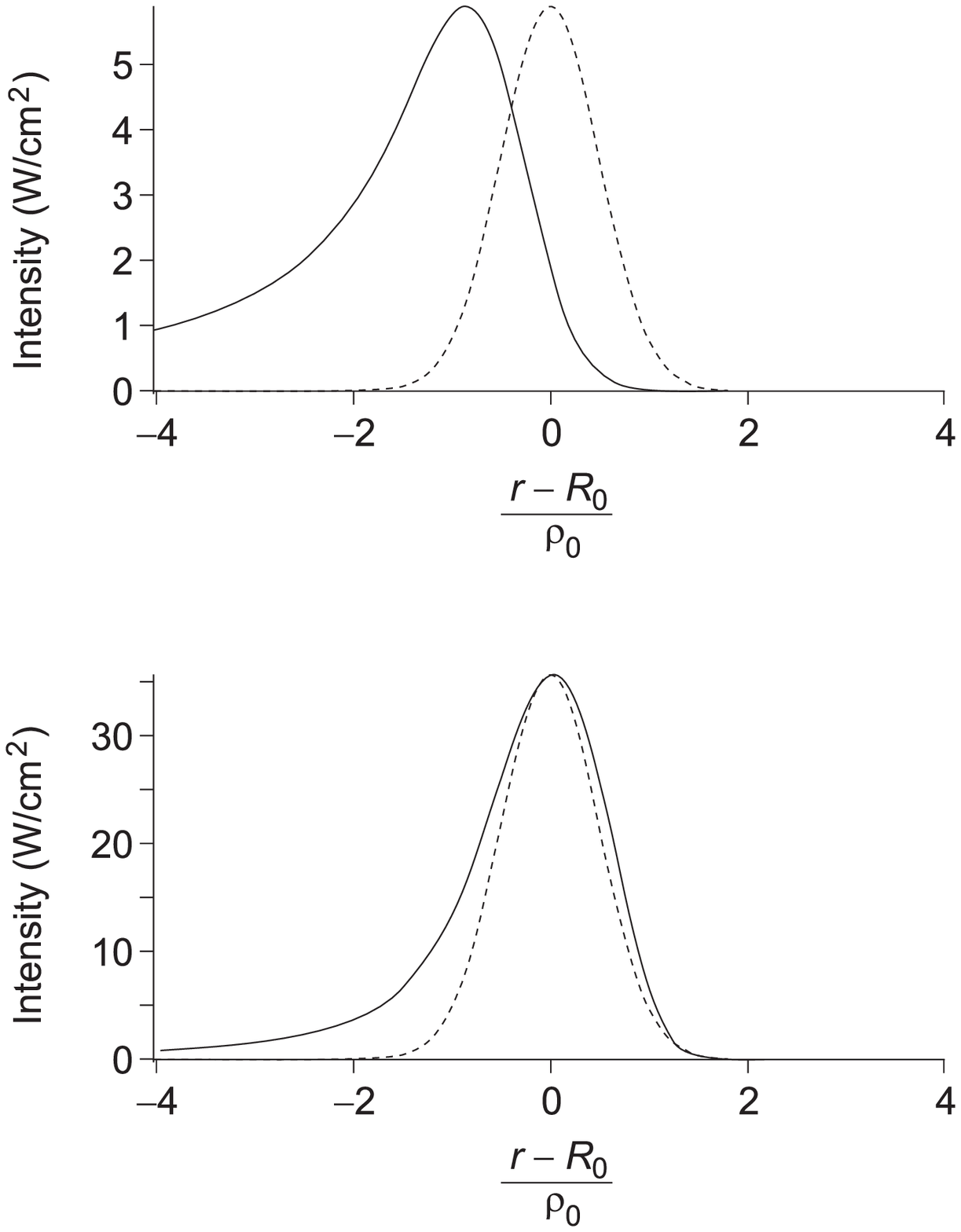}}
\caption{The full lines show the theoretical transverse intensity profile of
the beam in the focal plane of the convergent conical lens doublet for an
incident waist $w_i=645$~$\protect\mu$m in(a) and $w_i=3$~mm in (b). The
dashed lines gives the gaussian distributions with a width $\protect\rho_0$,
predicted in \protect\cite{Belanger78}. Vertical scales as in Fig. \ref%
{fig:diff}.}
\label{fig:compare}
\end{figure}

Fundamental differences appear between the present configuration and the
predictions of \cite{Belanger78}. In particular, the beam size in the focal
plane is almost $2.4$ times larger than expected (Fig. \ref{fig:compare}a).
If this broadening is effectively linked to the curvature of the conical
lens vertex, we expect that it decreases when the incident beam waist
increases. Fig. \ref{fig:compare}b shows that indeed, if $w_{i}$ is
increased significantly, the width of the ring becomes similar to that
predicted in \cite{Belanger78}.

\begin{figure}[tph]
\centerline{\includegraphics[width=10cm]{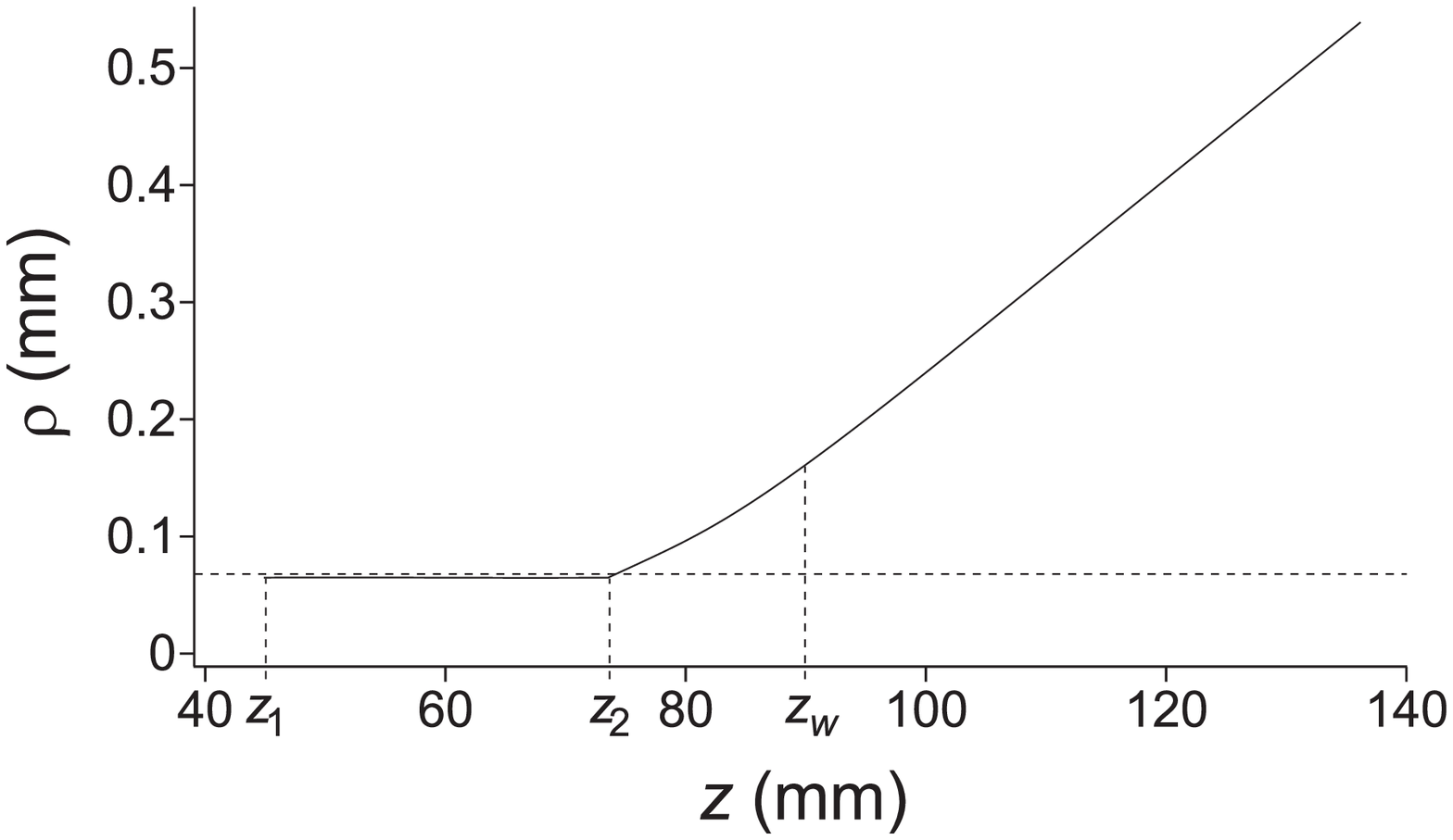}}
\caption{Intensity distribution thickness versus $z$. The width is measured
at $1/e^2$. The dashed line is the value predicted in the focus plane in %
\protect\cite{Belanger78}. $z_w$ is the abscissa of the focus plane.}
\label{fig:wvsz}
\end{figure}

Fig. \ref{fig:wvsz} shows how the ring thickness, measured at $1/e^{2}$ of
its maximum intensity, evolves as a function of $z$. Typically, three zones
can be defined, delimited by the values $z_{1}\simeq 45%
\mathop{\rm mm}%
\simeq 1.9\,z_{0}$ and $z_{2}\simeq 75%
\mathop{\rm mm}%
\simeq 3.1\,z_{0}$. For $z<z_{1}$, the main peak of the distribution is in $%
r=0$, as it is typical for the Bessel zone. For $z_{1}<z<z_{2}$, the
thickness is almost constant, with $\rho =1.6w_{0}\simeq \rho _{0}$, where $%
w_{0}$ is the minimum waist of the single convergent lens, in the focus
plane, and $\rho _{0}$ is the value predicted in \cite{Belanger78}. In the
third zone, referred as expanding zone in the following, and corresponding
to $z>z_{2}$, $\rho $ grows almost linearly with $z$. This behavior is very
different from that expected. In particular, the narrowest intensity
distribution is not in the focus plane, but in the constant zone, just after
the Bessel area. However, the smaller ring thickness corresponds to that
predicted by \cite{Belanger78} in the focus plane, although it occurs
outside the focus plane.

\begin{figure}[tph]
\centerline{\includegraphics[width=10cm]{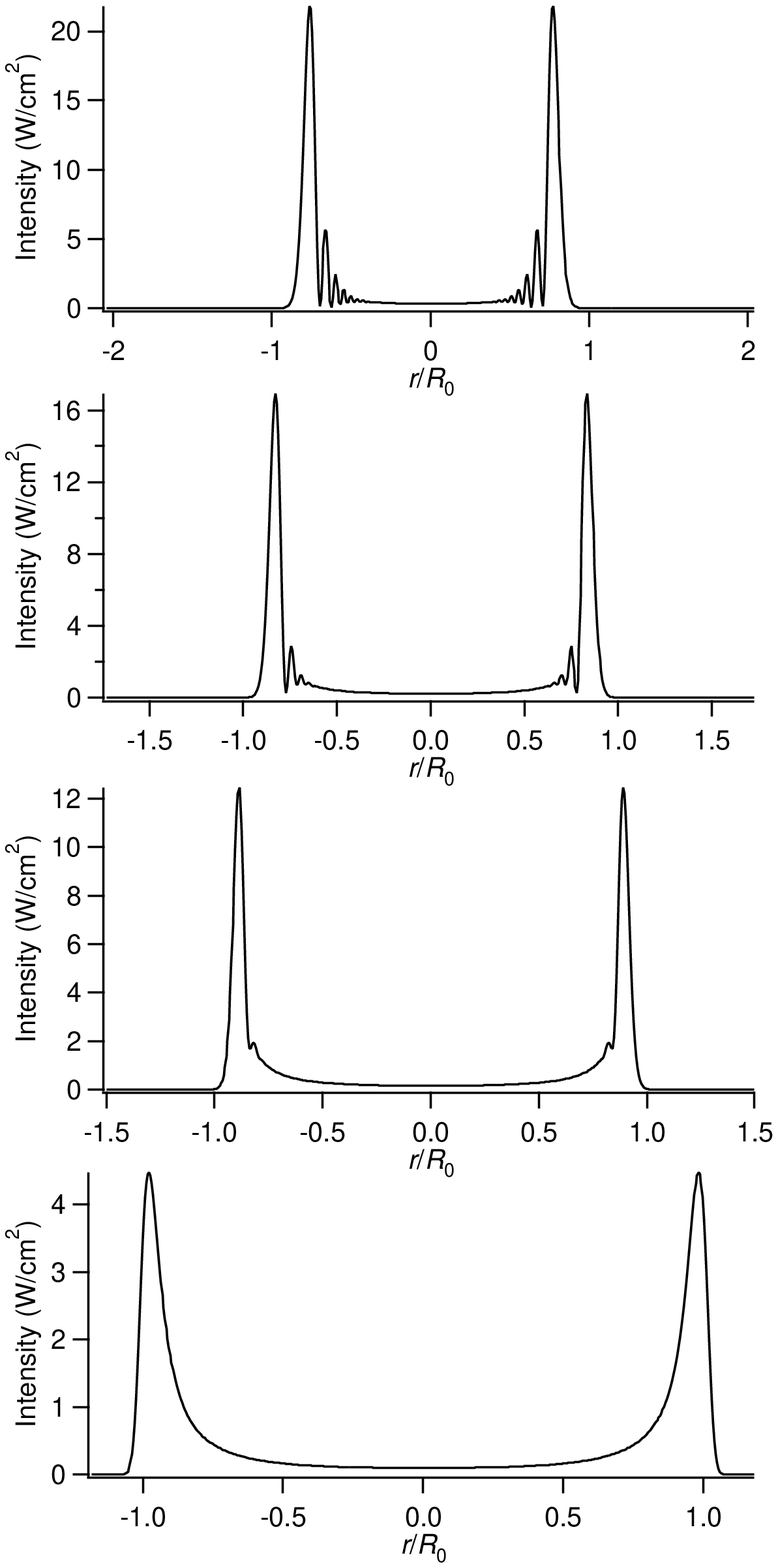}}
\caption{Theoretical transverse intensity profile of the beam after the
convergent conical lens doublet for the same parameters as in Fig. \ref%
{fig:fitdoublet} for (a) $z=55\mathop{\rm mm}=2.3z_0$, (b) $z=65 
\mathop{\rm
mm}=2.7z_0$, (c) $z=75\mathop{\rm mm}=3.1z_0$ and (d) $z=95 
\mathop{\rm mm}=4.0z_0$. All $r$ axes have the same absolute scale, from $%
r=0 \mathop{\rm mm} $ to $r=2\mathop{\rm mm}$. Vertical scales as in Fig. %
\ref{fig:diff}}
\label{fig:distrib}
\end{figure}

But the main novelties of our model concern the shape of the transverse
distribution of the intensity. As shown in Fig. \ref{fig:distrib}, very
different shapes are obtained in the constant and expanding zone. In the
constant zone, the external narrow ring is accompanied by several secondary
rings located inside the main one, and separated by dark rings. In the
immediate proximity of the Bessel zone, the minimum of the dark rings is
almost zero (Fig. \ref{fig:distrib}a), and it increases as the expanding
zone is approach (Fig. \ref{fig:distrib}b). At the edge of the constant and
expanding zones (Fig. \ref{fig:distrib}c), the contrast between secondary
and dark rings vanish, so that in the expanding zone, the transverse
distribution may be described by one single asymmetric ring with a tail
towards its center. The resulting pattern in the expanding zone seems to be
more clean, but it requires the use of a mask to suppress the intensity at
the center of the ring, leading to new diffracting distortions of the
pattern. On the contrary, in the constant zone, and in particular in the
proximity of the Bessel zone (in $z_{1}$), a mask adjusted to the first dark
ring cleans the pattern to the single external ring, without additional
distortion. The resulting pattern is a single narrow ring with two stiff
sides. Thus the constant zone appears to be the most appropriate to produce
clean hollow beams.

Another important aspect in the experimental realization of hollow beams is
either the total light power inside the final beam, or the peak intensity of
the ring. The first point depicts the losses introduced by the mask, while
the second one is linked to the thickness of the ring. If the mask is placed
at its optimal location $z_{1}$, as described above, it introduces losses of
22\%, corresponding to the part of power distributed in the secondary rings.
This value have to be compared to that obtained when the mask is placed in
the focus plane, as suggested in \cite{Belanger78}. In this case, the losses
depend naturally on the mask radius, and thus of the ring thickness. To
obtain a thickness comparable to that in $z_{1}$, we introduce 40\% of
losses. To reduce losses to 22\%, it is necessary to increase the thickness
by a factor 2.3: it is clear that the use of a mask in $z_{1}$ produces
narrower rings with less losses. As a consequence, as it can be easily
realized by comparing Fig. \ref{fig:compare}a and Fig. \ref{fig:distrib}a,
the peak intensity of the ring in almost four times larger in $z_{1}$ than
in $z_{w}$. So from the energetic point of view also, the technique
described above leads to a substantial gain.

\begin{figure}[tph]
\centerline{\includegraphics[width=10cm]{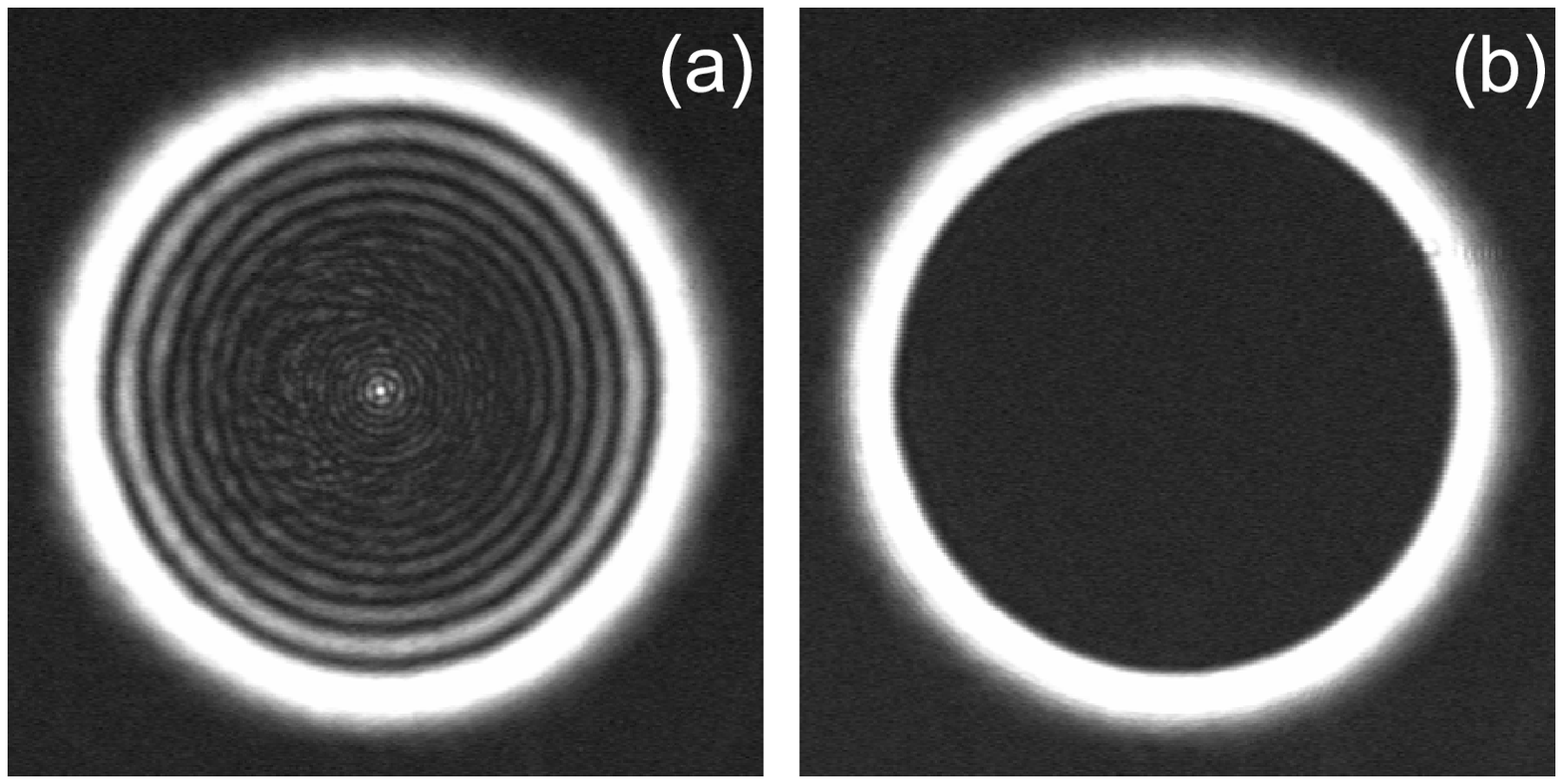}}
\caption{Experimental record of the intensity pattern obtained in $z_1$ for
the converging conical lens, with $f=500\mathop{\rm mm}$ and $d=20 
\mathop{\rm mm}$. In (a), without the mask; in (b), with a 2 mm mask.}
\label{fig:pattern}
\end{figure}

In the whole preceding discussion, we have emphasized one given
configuration, with a lens with focal length $f=100%
\mathop{\rm mm}%
$, with a distance $d=10%
\mathop{\rm mm}%
$ between the lens and the axicon. The conclusions apply of course to other
values of $f$ and $d$. As an example, we illustrate in Fig. \ref{fig:pattern}
the hollow beam obtained for a different set of parameters, using the above
technique. The resulting pattern is a single ring with a stiff inner border,
and losses introduced by the mask are measured to be $0.19\pm 0.01$, in good
agreement with the theory.

\section{Conclusion}

We have shown that taking into account the curvature of the
conical lens vertex leads to an accurate description of the beam
transformation through a conical lens and a convergent conical lens doublet.
Calculations are in excellent agreement with the experiments. We show that
the use of a mask allows to obtain an excellent quality hollow beam,
assuming that the mask is located adequately, just after the Bessel zone.
The quality of the hollow beam consists in a single regular ring, with an
intensity going abruptly to zero on its inner side, and no residual
intensity in the center of the ring. This enables the use of such beams in
interferometric experiments, as e.g. in the realization of optical lattices%
\cite{nous02}. In most of the applications, the present system will win in
convenience if a second conical lens, located just after the Bessel zone of
the first one, is added to the convergent conical lens doublet. Indeed, the
second axicon will rectify the phase surfaces and fix the ring radius versus
propagation. A third conical lens before the doublet may be added to control 
independently $R_0$ and $\rho$\cite{Song99}.

The Laboratoire de Physique des Lasers, Atomes et Mol\'{e}cules is ``Unit%
\'{e} Mixte de Recherche de l'Universit\'{e} de Lille 1 et du CNRS'' (UMR
8523). The Centre d'Etudes et de Recherches Lasers et Applications (CERLA)
is supported by the Minist\`{e}re charg\'{e} de la Recherche, the R\'{e}gion
Nord-Pas de Calais and the Fonds Europ\'{e}en de D\'{e}veloppement
Economique des R\'{e}gions.


\end{document}